\documentclass[prodmode,acmtomccap]{acmlarge2}
\usepackage{balance}
\usepackage{setspace2}

\usepackage{amsmath,enumerate,mathtools}
\usepackage{enumitem,placeins}
\usepackage[low-sup]{subdepth}
\usepackage{fp}
\usepackage{physymb}
\usepackage{graphicx}
\usepackage{bm}
\usepackage{booktabs}
\usepackage{tikz}
\usepackage{amsmath}
\usepackage{amssymb}
\usepackage{amsthm}
\usepackage{latexsym}
\usepackage{verbatim}
\usepackage{graphicx}
{\begin{list}               %    with flush left bullets.
    {$\bullet$ \hfill}{
        \setlength{\leftmargin}{\parindent}
        \setlength{\parsep}{0.04\baselineskip}
        \setlength{\itemsep}{0.5\parsep}
        \setlength{\labelwidth}{\leftmargin}
        \setlength{\labelsep}{0em}}
    }
{\end{list}}

\providecommand{\cref}[1]{Chapter~\ref{chap:#1}}

\providecommand{\fref}[1]{Fig.~\ref{fig:#1}}

\providecommand{\tref}[1]{Table~\ref{tab:#1}}

\providecommand{\R}{\ensuremath{\mathbb{R}}}
\providecommand{\C}{\ensuremath{\mathbb{C}}}

\DeclareMathOperator*{\argmin}{\arg\!\min}
\providecommand{\E}[1]{\mathbb{E}\left[#1\right] }

\providecommand{\He}{\mathrm{H}}

\DeclareMathOperator{\Tr}{Trace}
\providecommand{\diag}{\mathop{\mathrm{diag}}}

\renewcommand{\vec}[1]{\ensuremath{\mathbf{#1}}}
\providecommand{\mat}[1]{\ensuremath{\mathbf{#1}}}

\providecommand{\mA}{\mat{A}}

\providecommand{\mI}{\mat{I}}

 \providecommand{\mR}{\mat{R}}

\providecommand{\mW}{\mat{W}}

\providecommand{\mX}{\mat{X}}

\providecommand{\va}{\vec{a}}

 \providecommand{\vn}{\vec{n}} 
 \providecommand{\vp}{\vec{p}}
 \providecommand{\vr}{\vec{r}}
\providecommand{\vs}{\vec{s}} 
 \providecommand{\vw}{\vec{w}}

\providecommand{\vx}{\vec{x}} \providecommand{\vy}{\vec{y}}

\graphicspath{ {../figures/} }

\acmVolume{2}
\acmNumber{3}
\acmArticle{0}
\acmYear{2010}
\acmMonth{5}

\title{Collaborative Randomized Beamforming for Phased Array Radio Interferometers} 
\author{ORHAN \"{O}\c{C}AL, PAUL HURLEY, GIOVANNI CHERUBINI and SANAZ KAZEMI \affil{IBM Zurich Research Laboratory, CH-8803 R\"{u}schlikon, Switzerland}}

\begin{abstract}
The Square Kilometre Array (SKA) will form the largest radio telescope ever
built and such a huge instrument in the desert poses enormous engineering
and logistic challenges. Algorithmic and architectural breakthroughs are needed.

Data is collected and processed in groups of antennas before
transport for central processing. This processing includes beamforming,
primarily so as to reduce the amount of data sent. The principal existing
technique points to a region of interest independently of the sky
model and how the other stations beamform.

We propose a new collaborative beamforming algorithm in order to maximize
information captured at the stations (thus reducing the amount of data
transported). The method increases the diversity in measurements through
randomized beamforming. We demonstrate through numerical simulation the
effectiveness of the method. In particular, we show that randomized beamforming
can achieve the same image quality while producing $40\%$ less data when
compared to the prevailing method matched beamforming.
\end{abstract}

\category{Beamforming, array signal processing, interferometry, radio astronomy, sparse signal processing}{}{}
\begin{document}

\maketitle

\section{Introduction}
\label{sec:intro}

The Square Kilometre Array (SKA) will form, upon completion, the largest and the
most sensitive radio telescope ever built, consisting of millions of antennas
over a total collection area of one square kilometer~\cite{Dewdney:2009bq}. The
resultant data will be immense -- on the order of one terabyte of data every
second -- equivalent to almost one tenth the total global internet traffic. The
state of the art in engineering and algorithms for data collection, instrument
calibration, storage, and imaging struggles to keep pace.

In addition to hardware optimization, tailored improved algorithms with lower
data production are a promising solution. Thus, the work presently described was
motivated by the need to reduce data as far up the processing chain as possible.

SKA will comprise different antenna types, dishes and phased arrays. The phased
array is designed to attain a large field of view. Antennas are grouped into
\emph{antenna stations} where RF signals are received. These signals are then
processed by beamforming, and sent to a central data processor for image
creation through correlation and further beamforming~\cite{Wright:2002th}.

%The introduction of hierarchically designed radio telescopes with phased array
%stations is relatively new
%\cite{VanHaarlem:2013vf,Lonsdale:2009ij,Ellingson:2009gg}. 
For station beamforming, methods from array signal processing can be used to
optimize different criteria. The strategy currently used in one SKA pathfinder,
the Low Frequency Array (LOFAR) \cite{VanHaarlem:2013vf}, is \emph{matched
beamforming} \cite{vanderVeen:2013uz}. Antenna RF signals are summed after phase
aligning the signal coming from an a priori chosen look direction. This can be
seen as a matched spatial filter \cite{Wijnholds:2010wn}, where the filter is
matched towards the chosen direction. Another method used is minimum variance
directionless response (MVDR) beamformer, equivalent to maximizing the ratio of
signal power from a chosen direction to interference plus
noise~\cite{vanderVeen:2013uz,Capon:1969tf}.

% It is reported that the MVDR
% beamformer performance degrades substantially with errors in array parameters,
% such as antenna positions and antenna gains \cite{Lorenz:2005tx}. Hence, robust
% MVDR beamformer methods that account for errors in array response and
% uncertainties in the look direction are also proposed
% \cite{Lorenz:2005tx,Beck:2007ub}.

It is neither feasible nor desirable to send raw antenna time-series for
correlation. Beamforming, mapping down signals from a higher to a lower
dimensional space, is essentially lossy data compression distributed throughout
the stations. Looked at it from that viewpoint, the following question naturally
arises: how could one maximize the information content so as to reduce the data
transmitted from stations to the central processor?

%Subsequent questions then follow. By freeing up beamforming, could we render the
%instrument more flexible for the many different investigations and science
%cases, both envisaged and to be envisaged over the instrument's lifetime?

Thus, our goal is to reduce the amount of data and the complexity of the
subsequent stages. Less data coming out of the stations means less traffic sent to
 the central data processor. Hence, an early reduction in the data
yields savings in data transportation cost and the amount of processing in the later stages.

% The organization of the paper is as follows. Section~\ref{sec:sig_proc} states
% the signal model. Section~\ref{sec:imaging_methods} covers image generation
% methods from interferometer measurements, which are going to be used in
% describing reconstruction of the sky image by the proposed beamforming methods.
% Section~\ref{sec:background} describes the problem of beamforming at stations
% and where the proposed algorithms apply. We then propose in
% Section~\ref{sec:col-beam} a collaborative beamforming technique for reducing
% data rates. Section~\ref{sec:sim} provides a numerical evaluation of the
% methods, before conclusion in Section~\ref{sec:conclusions}.

The organization of the paper is as follows. Section~\ref{sec:background} describes beamforming at stations,
including a description of
the signal model. We then propose in
Section~\ref{sec:col-beam} a collaborative beamforming technique for reducing
data rates, and a sparse signal recovery formulation leveraging the
proposed beamforming. Section~\ref{sec:sim} provides a numerical
evaluation before conclusion in Section~\ref{sec:conclusions}.

\section{Beamforming at stations}
\label{sec:background}

\subsection{Signal Model}
\label{sec:sig_proc}

%\begin{figure}[t]
%	\centering
%	\includegraphics[width=0.6\linewidth]{far_field-crop}
%	\caption{
%Under far field assumption, signal coming from any direction $\vr$ reaches the
%antennas in parallel.
%		}
%	\label{fig:far_field}
%\end{figure}

Let $L$ denote the number of antennas and $Q$ the number of sources. Assume that
celestial sources are in the far field, signals emitted by them are narrow band
circularly-symmetric complex Gaussian processes, and that signals emanating from
different directions in the sky are uncorrelated~\cite{Taylor:1999vv}. The
signal received by the $L$ antennas, $\vx(t): \R \rightarrow \C^L$, can be
stated as
\begin{align*}
	\vx(t) = \sum_{q=1}^Q \va_q s_q(t) + \vn(t),
\end{align*}
where $s_q(t) \sim \mathcal{NC}(0,\sigma^2_q)$ is the signal emitted by source
$q$, $\vn(t) \sim \mathcal{NC}(0,\sigma^2 \mI)$ is the additive noise at the
antennas, and $\va_q \in \C^L$ is the \emph{antenna steering vector} towards
source $q$ given by~\cite{vanderVeen:2013uz}
\begin{align}
	\va_q =
	\begin{pmatrix}
		e^{-j \frac{2\pi}{\lambda} \vr_q^\top \vp_1} &
		e^{-j \frac{2\pi}{\lambda} \vr_q^\top \vp_2} &
		\cdots &
		e^{-j \frac{2\pi}{\lambda} \vr_q^\top \vp_L}
	\end{pmatrix}^\top,
	\label{eq:antenna-steering-vector}
\end{align}
where $\lambda$ is the observation wavelength, $\vr_q$ is the unit vector
pointing at source $q$, and $\vp_1$, $\cdots$ $\vp_L$ are the positions of the
antennas. This summation can be written as a matrix vector product
\begin{align*}
	\vx(t) = \mA \vs(t) + \vn(t),
\end{align*}
where $\mA \in \C^{L \times Q}$ has its column $q$ equal to $\va_q$, and
$\vs(t): \R \rightarrow \C^Q $ has its $q$th element equal to $s_q(t)$.

In a wide number of scientific cases, the parameters of
interest are the source intensities (signal variance) and
positions. Because signals arriving at antennas are modeled as circularly
symmetric complex Gaussian processes, the autocorrelation matrix is a sufficient
statistic. Assuming the noise and the signals to be uncorrelated, we have the
correlation matrix
\begin{equation}
	\mR = \E{ \vx(t) \vx(t)^\He }
% 	&= \E{ \left( \mA \vs(t) + \vn(t) \right) \left( \mA \vs(t) + \vn(t) \right)^\He } \notag \\
	= \mA \vec{\Sigma_s} \mA^{\He} + \vec{\Sigma_n},
	\label{eq:array-signa-processing-representation}
\end{equation}
where $\vec{\Sigma_s}=\diag(\sigma^2_1,\sigma^2,\ldots,\sigma^2_Q)$ and $\vec{\Sigma_n}= \sigma^2_n \vec{I},
$ are diagonal covariance matrices for
the signal and noise respectively.
%as
%\begin{align}
%	\vec{\Sigma_s} &= 
%	\begin{pmatrix}
%		\sigma^2_1 &  & 0 \\
%		  & \ddots &  \\
%		0 & & \sigma^2_Q
%	\end{pmatrix}, \\
%	\vec{\Sigma_n} &= \sigma^2_n \vec{I}_L,
%	\label{eq:cov-matrices}
%\end{align}
%where $\vec{I}_L$ is the $L \times L$ identity matrix.

\subsection{Beamforming}
Here we lay the groundwork for explaining the beamforming methods described in
Section \ref{sec:col-beam}.
% In particular, we will discuss how stations, groups
% of antennas, play the role of primary receiving elements, and how
% station beamforms play the role of primary beam shapes. 
In particular, we will describe the hierarchical design of phased array radio
interferometers, and then discuss the general issue of combining signals at
station level and the current prevailing method.

\begin{figure}[t]
	\centering
	\includegraphics[width=0.7\linewidth]{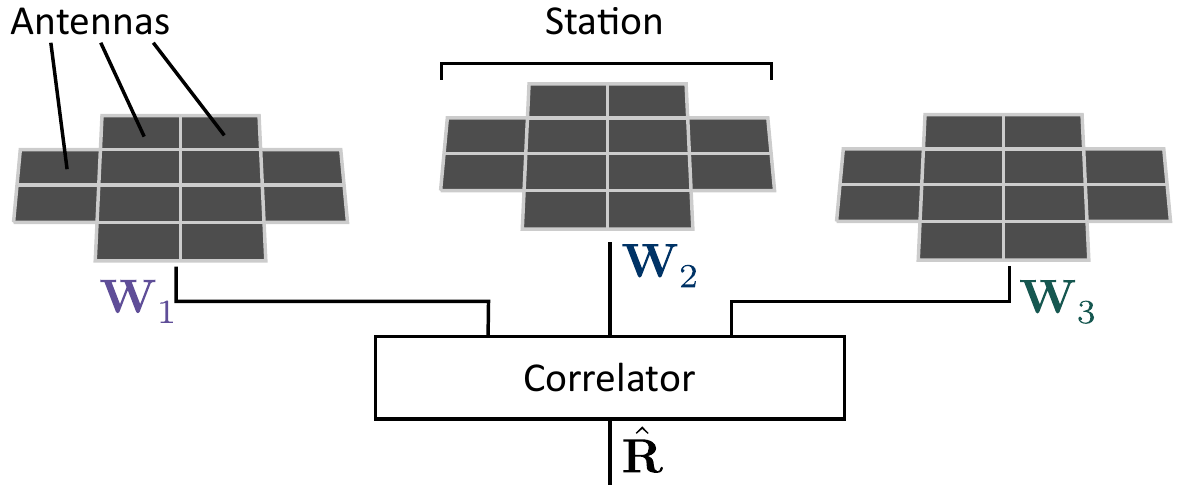}
	\caption{
Hierarchical design of phased array radio telescopes. Groups of antennas are
combined as stations.}
	\label{fig:bf}
\end{figure}

In hierarchically designed phased array radio
interferometers~\cite{VanHaarlem:2013vf,Ellingson:2009gg,Lonsdale:2009ij},
multiple antennas are grouped according to geography into \emph{antenna
stations}, cf. \fref{bf}. The positions and the orientations of the antennas are
fixed.
% and they cannot be physically oriented towards the regions of interest of the sky. 
In order to scan different portions of the sky, individual antennas
have a large field of view~\cite{Wijnholds:2010wn}, cf. \fref{wide_beams}. Sending
all data received from all antennas for central data processing is costly.
Hence, the data is typically reduced at stations by beamforming.

\begin{figure}[b]
	\centering
	\includegraphics[width=0.5\linewidth]{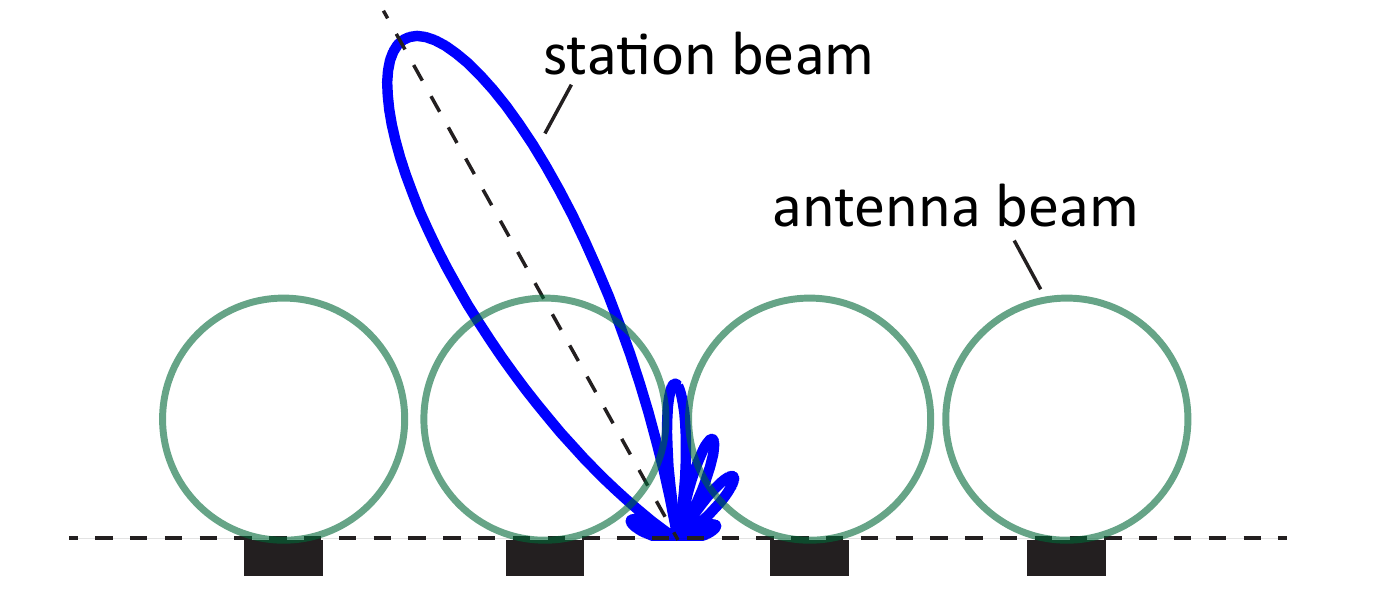}
	\caption{
Antenna beams capture signals from a large region (green circles). Matched
beamforming reduces the field of view and amplifies the signal coming from a
particular direction.}
	\label{fig:wide_beams}
\end{figure}

The current prevailing method is \emph{matched beamforming}, which points
towards the center of a chosen region. Formally, say we choose a region centered
around direction $\vr_0$. Then, the signals are combined using a beamforming
vector equal to the antenna steering vector
%(cf.\eqref{eq:antenna-steering-vector}) 
towards $\vr_0$. If all of the stations have the same layout -- which is the
case in modern radio interferometers, e.g., LOFAR~\cite{VanHaarlem:2013vf} --
then all beamforming vectors are the same, leading to the same beam shape at each
station.

One approach taken to introduce variations in beam shapes is to rotate
the stations with respect to each other~\cite{VanHaarlem:2013vf}. These
rotations may help to reduce 
%systematic problems due to the beam-shapes, in particular
grating lobes,
%If the minimum spacing between
%antennas within a station is larger than half the wavelength of observation,
%the resulting beam has areas outside the chosen region of interest that will be
%weighted as strongly as the sources within that region.
%, cf.\fref{grating-lobes}. 
whose positions are also rotated, and hence the signals outside
the region of interest are averaged out.
%By rotating the stations with respect to each other, the positions of the
%grating lobes of the stations are also rotated, hence the signals from outside
%the region of interest are averaged out.

%\begin{figure}[b]
%	\centering
%	\includegraphics[width=0.8\linewidth]{grating-lobes-2}
%	\caption{
%Grating lobes due to subsampling in the low frequencies. The array has $4$
%antennas equally spaced on a line with $0.8$ wavelengths, and the beamforming is
%done by conjugate matching towards $30$ degrees. Plot shows the magnitude
%scaling towards the directions.}
%	\label{fig:grating-lobes}
%\end{figure}

%In its general form, 
Station beamforming can be viewed as linear operation on a $J$ dimensional
signal, where $J$ is the number of antennas at the station. It is thus
representable, for $M$ beams at the $k$th station, by matrix $\mW_k \in \C^{J
\times M }$. The beamformer output is then
\begin{align*}
	\vy_{k}(t) 
% 	&= \mW^{(k)\He} \vx^{(k)}[n] \\
	= \mW_{k}^{\He} \left( \mA_{k} \vs(t) + \vn(t) \right) ,
\end{align*}
where $(\cdot)^\He$ denotes conjugate transpose.
% of a vector. 
The correlation of two beamformed outputs is
% equal to
\begin{equation}
	\E{ \vy \vy^{\He} } = \mW^\He \mA \vec{\Sigma}_s \mA^\He \mW + \mW^\He \vec{\Sigma}_n \mW,
	\label{eq:beamformer-signal}
\end{equation}
where $\mW \in \C^{JL \times ML} $ is the block diagonal matrix containing the
beamforming matrices of $L$ stations, and $\mA \in \C^{JL \times Q} $ the
response matrix of all antennas towards the $Q$ sources.

Let us study the output from a station $k$. Dropping the dependence on time for
brevity, the signals at the antennas can be written as $\vx_{k} = \mA_{k} \vs +
\vn_{k}$. Then $\vy_{k} = \vw_{k}^{\He} \left( \mA_{k} \vs + \vn_{k} \right)$
for beamforming vector $\vw_{k}$. The variance of the resulting signal is equal
to
\begin{align*}
	\E{\vy_{k} \vy_{k}^{\He} } 
%	&= \vw^{(k)\He} \E{\vx^{(k)\He}\vx^{(k)} } \vw^{(k)} \\
% 	&= \vw^{(k)\He} \left( \sum_{q=1}^Q \va_q^{(k)} \va_q^{(k)\He} \sigma^2_q + \sigma^2_n \mI  \right) \vw^{(k)} \\
	&= \sum_{q=1}^Q \vert \vw_{k}^{\He} \va_{k,q} \vert^2 \sigma^2_q + \sigma^2_n \Vert \vw_{k} \Vert^2,
\end{align*}
where $\va_{k,q}$ stands for column $q$ of $A_k$. Hence, the variance of the
signal from the $q$th source in the beamformer output of the $k$th station,
$\sigma_{k,q}^2$, satisfies
\begin{align}
	\sigma_{k,q}^2 &= \big\vert \vw_k^{\He } \va_{k,q} \big \vert^2 \sigma^2_q 
	\overset{(a)}{\leq} \big\Vert \vw_{k} \big\Vert^2 \big \Vert \va_{k,q} \big \Vert^2 \sigma^2_q,
	\label{eq:var-of-direction}
\end{align}
where $(a)$ follows from Cauchy-Schwartz inequality. If we use the beamforming
vector  $\vw_{k} = \va_{k,q}
/ \sqrt{J} $,  a multiple of the antenna steering vector, then $(a)$ becomes an equality. This is the mathematical description of
matched beamfoming. In that case, $\sigma^2_{k,q} = J \sigma^2_q$. Because this
beamforming vector is of unit norm, the noise variance at beamformer output is
equal to the variance $\sigma^2_n$ at a single antenna. Hence, the signal coming
from the $q$th source is amplified with a factor of $J$ with respect to the
noise.

% An intuitive explanation behind matched beamforming is, under the assumption of
% narrowband signals, the signal coming from a source at a certain direction
% reaches the antennas with only a phase difference. By matched beamforming, the
% signal that comes from a direction is phase-aligned and summed. Hence, the
% magnitude of the signal is maximized. On the other hand, because noise is
% uncorrelated at different antennas, it is not coherently combined.

Intuitively, narrowband signal coming
from a particular direction reaches the antennas with phase delays. Matched
beamforming aligns the phases of signals from a chosen direction and sums.
Hence, the magnitude of the signal is maximized, whereas because noise is
uncorrelated at antennas, it is not coherently combined.

\section{Collaborative beamforming}
\label{sec:col-beam}

%\begin{figure}[b]
%	\centering
%	\begin{subfigure}[b]{0.48\linewidth}
%		\centering
%		\includegraphics[width=\textwidth]{phase_coding}
%		\caption{}
%		\label{fig:}
%	\end{subfigure}
%	\begin{subfigure}[b]{0.48\linewidth}
%		\centering
%		\includegraphics[width=\textwidth]{amplitude_coding}
%		\caption{}
%		\label{fig:}
%	\end{subfigure}
%	\caption{
%(A) Using the same beam shape scales signals equally, there is no positions
%information in the magnitude. (B) Different beam shapes gives varying weights to
%signals coming from different directions. Magnitudes can be leveraged for
%estimating the angle of arrival.}
%	\label{fig:phase-coding}
%\end{figure}

We have seen that beamforming is essentially fixed/static at all stations. This
is a waste. When all beam shapes are the same, we cannot use the magnitude
information in sky image reconstruction, but are limited to the phase
information resulting from geometric delays between stations. When stations use
different beamforms,
% the correlation between the beamformed outputs of two antenna stations becomes
the correlation between two station outputs becomes
\begin{align}
	\E{ \vy_{k} \vy_{m}^{\He}  } &= \sum_{q=1}^Q \vw_{k}^{\He} \va_{k,q} \va_{m,q}^{\He}  \vw_{m}  \sigma^2_q.
	\label{eq:correlation-between-two-beamfors}
\end{align}
As  $\vw_{k}^{\He}
\va_{k,q} \va_{m,q}^{\He} \vw_{m} \in \C$, the correlation equals the sum of signal powers weighted by a constant
that depends on station indices
$k$ and $m$. Because of scaling variations,  information is contained not only in the phases but also 
the signal magnitudes. This increase in information can then be
exploited in non-linear signal recovery methods for the improvement of the quality
of the resulting image.

Another benefit of using different beamforms is a reduction in the effects of
potential systematic errors, such as grating lobes that result from fixed beam
shapes.

\subsection{Randomized beamforming}
We can reduce the amount of data required for the same image quality (or conversely
increase image quality for the same data rate). This is vital to make an SKA in
the middle of the desert feasible.  To get different beam shapes to act collaboratively for 
this purpose, beamforming vectors at antenna stations can be chosen from a random
distribution. This method is inspired by compressed sensing, where random
measurements were shown to preserve the information contained in sparse signals
\cite{Donoho:2006vp}.

%The beam shape of a station defines how each signal's power is scaled. The
%correlation between beamformer outputs of two different stations multiplies
%their individual scalings. If there are $L$ station, then we get $L^2$ many
%different scalings for the same source, in addition to the phase information
%resulting from the baselines between the stations.

The multiple beams at station $k$ can be described by a beamforming matrix
$\mW_{k}$ (cf. Section \ref{sec:col-beam}), which can be chosen with particular
goals in mind:
\begin{enumerate}[label=R\arabic{enumi}:,ref=R\arabic{enumi}]
  \item 
generate each matrix element from an independent and identically distributed
circularly-symmetric complex Gaussian random distribution with unit variance,
and then normalize the columns to be of unit norm;
\label{en:R1}
	\item
after generating the matrix as in \ref{en:R1}, convolve each column with a fixed
beam shaping filter and truncate the excess length (this attenuates signals
coming from outside a chosen region of interest); or \label{en:R2}
	\item 
generate the matrix by matched beamforming towards randomly chosen directions
within the chosen region of interest.
\label{en:R3}
\end{enumerate} 
\fref{eb1} to \fref{eb4} illustrate typical beams from these strategies.
\ref{en:R1} has the largest beam variation. Vector normalization ensures the
noise variance at beamformer output is equal to the antenna noise variance.
However, depending on beam shape, we may attenuate the signal coming from a
source as evident by \eqref{eq:var-of-direction}. Hence, there is a trade off
between beamform variations and signal power. When only interested in a small
part of the sky in the presence of high measurement noise, using
strategy~\ref{en:R2}
% where the beam
%shaping filter equal to the matched beamformer towards the region of interest,
or~\ref{en:R3} would be more promising.

\begin{figure}[tp]
    \begin{minipage}[t]{0.45\linewidth}
        \centering
	\includegraphics[clip=true,trim=0 20 0 0, width=\textwidth]{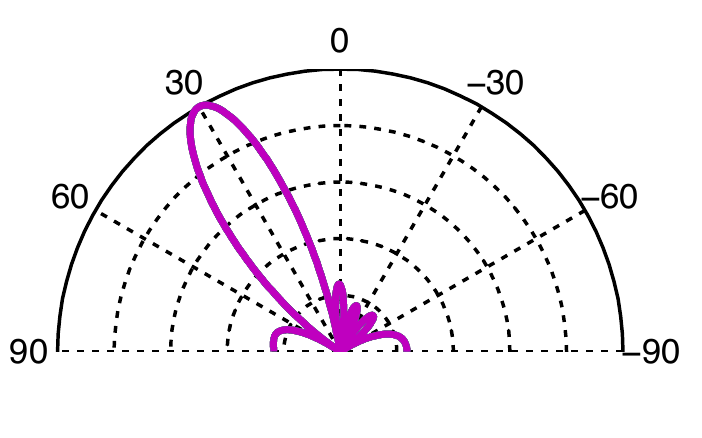}
	\caption{Matched beamforming}
	\label{fig:eb1}
    \end{minipage}
    \hspace{0.1\linewidth}
    \begin{minipage}[t]{0.45\linewidth}
        \centering
        \includegraphics[clip=true,trim=0 20 0 0, width=\textwidth]{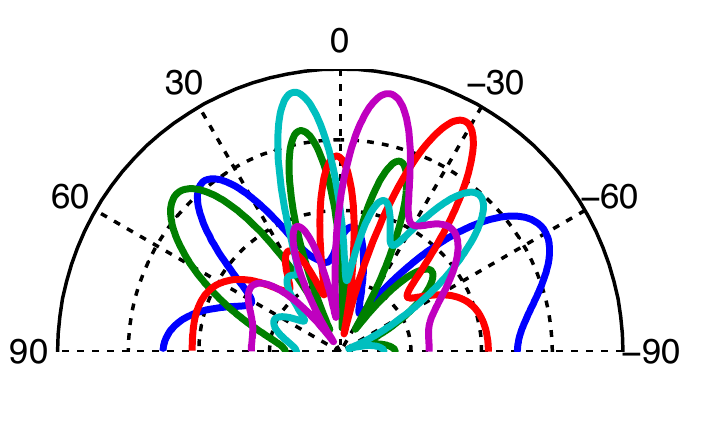}
	\caption{Randomization \ref{en:R1} }
	\label{fig:eb2}
    \end{minipage}
    
    \begin{minipage}[t]{0.45\linewidth}
        \centering
	\includegraphics[clip=true,trim=0 20 0 0, width=\textwidth]{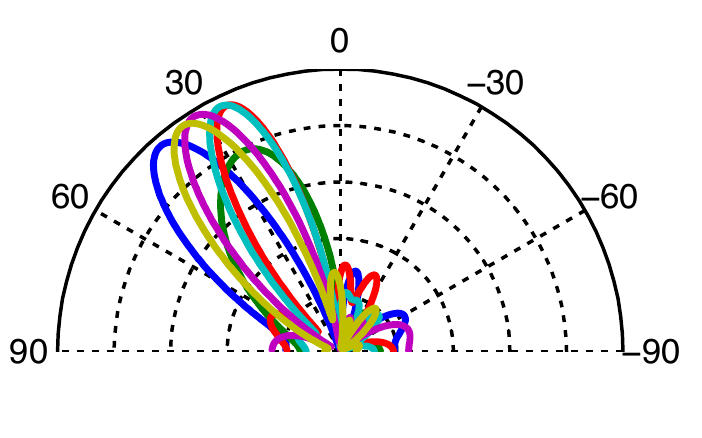}
	\caption{Randomization \ref{en:R2} }
	\label{fig:eb3}
    \end{minipage}
    \hspace{0.1\linewidth}
    \begin{minipage}[t]{0.45\linewidth}
        \centering
        \includegraphics[clip=true,trim=0 20 0 0, width=\textwidth]{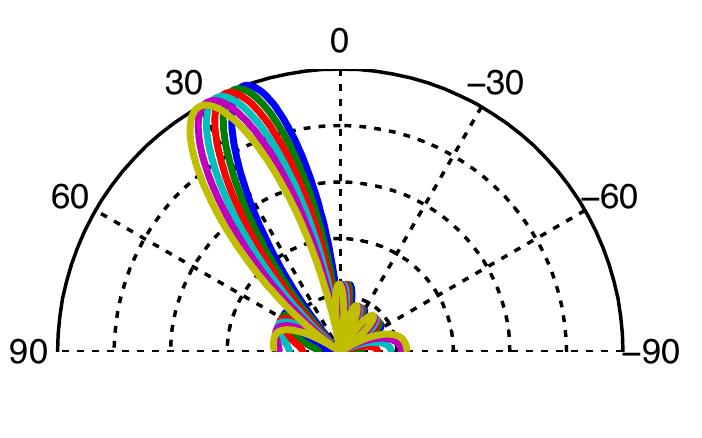}
	\caption{Randomization \ref{en:R3} }
	\label{fig:eb4}
    \end{minipage}
\end{figure}

% \begin{figure}%[t]
% 	\centering
% 	\begin{subfigure}[b]{0.49\linewidth}
% 		\centering
% 		\includegraphics[clip=true,trim=0 20 0 0, width=\textwidth]{beamforms-3}
% 		\caption{Matched beamforming}
% 	\end{subfigure}
% 	\begin{subfigure}[b]{0.49\linewidth}
% 		\centering
% 		\includegraphics[clip=true,trim=0 20 0 0, width=\textwidth]{beamforms-1}
% 		\caption{Randomization \ref{en:R1} }
% 	\end{subfigure}
% 
% 	\begin{subfigure}[b]{0.49\linewidth}
% 		\centering
% 		\includegraphics[clip=true,trim=0 20 0 0, width=\textwidth]{beamforms-rf}
% 		\caption{Randomization \ref{en:R2} }
% 	\end{subfigure}
% 	\begin{subfigure}[b]{0.49\linewidth}
% 		\centering
% 		\includegraphics[clip=true,trim=0 20 0 0, width=\textwidth]{beamforms-rd}
% 		\caption{Randomization \ref{en:R3} }
% 	\end{subfigure}
% 	\caption{
% Matched beamforming and realizations of randomized beamforming strategies
% \ref{en:R1}, \ref{en:R2} and \ref{en:R3}. Randomized beams can be used to
% collaboratively receive signals from a wide field of view, or increase
% measurement diversity from a region of interest.}
% 	\label{fig:example-beamforms}
% \end{figure}
% 
\subsection{Sparse signal recovery}
\label{sec:imaging_methods}

Randomized beamforming is best tailored to a model-based sparse reconstruction
algorithm. One such method is to maximize the likelihood of the observed data
with respect to the parameters of the model~\cite{vanderVeen:2013uz}. Radio
astronomy parameters are, for point source models, the number of sources,
positions and intensities of the sources, and the noise variance, which we
represent by the set $\theta = \{Q,\vr_1,\cdots, \vr_Q, \sigma^2_1, \cdots,
\sigma^2_Q, \sigma^2_n \}$.
When the samples $\{ \mX[n]\}_{n=1,\cdots,N} $ are independent the maximum
likelihood estimator is~~\cite{vanderVeen:2013uz}
\begin{equation}
	\argmin_{\theta} \,\, \log \det(\mR(\theta)) + \Tr \bigl( \mR^{-1}(\theta) \hat{\mR} \bigr).
	\label{eq:ml}
\end{equation}
%\begin{equation}
%	\begin{aligned}
%		& \underset{\theta}{\mathrm{\arg\!\min}}
%		& & \log \det(\mR(\theta)) + \Tr \left(  \mR^{-1}(\theta) \hat{\mR} \right).
%	\end{aligned}
%	\label{eq:ml}
%\end{equation}
where $\mR(\theta) = \mA(\theta) \vec{\Sigma(\theta)} \mA^{\He}(\theta) +
\sigma^2_n(\theta) \mI$. In the model, the position parameters show up in a
non-linear way, and the source intensities linearly. Given the source positions,
the noise variance and assuming $\mA$ admits a left inverse, the solution for
the source intensities is equal to the diagonal elements of $ (\mA^\He \mA)^{-1}
\mA^\He \hat{\mR} \mA (\mA^\He \mA)^{-1} - \sigma^2_n \mI$~\cite{Jaffer:1988iw}.

However, in the absence of a priori knowledge,
% of these, the positions of the sources, or if the number of sources is larger than the number of antennas, 
problem \eqref{eq:ml} is hard to solve. In such cases, because a point source
model imposes sparsity on the sky, sparse signal recovery methods, such as
algorithms from compressed sensing literature can be used. To this end, we
assume that the sources are on a two dimensional grid, denoted by
$\{\vr_i\}_{i=1,\cdots,N_g} $ where $N_g$ is the number of grid points. Define
the matrix $\mA_g \in \C^{L \times N_g} $ that has $k$th column equal to the
antenna steering vector towards the $k$th grid point. Then, the signal intensity
vector $\bm{\rho}_s \in \R^{N_g} $ can be estimated by least absolute shrinkage
and selection operator \cite{Tibshirani:1996wb} (LASSO) by
\begin{equation}
 \argmin_{\bm{\rho}_s ,\rho_n}
  \Vert \hat{\mR} - \mA_g \mathrm{diag} \left( \bm{\rho}_s \right) \mA_g^{\He} - \rho_n \mI \Vert_{F} + \lambda \Vert \bm{\rho}_s \Vert_{1},
	\label{eq:LASSO}
\end{equation}
%\begin{equation}
%	\begin{aligned}
%		& \underset{\bm{\rho}_s ,\rho_n }{\mathrm{minimize}}	
%		& &  \Vert \hat{\mR} - \mA_g \mathrm{diag} \left( \bm{\rho}_s \right) \mA_g^{\He} - \rho_n \mI_L \Vert_{F} + \lambda \Vert \bm{\rho}_s \Vert_{1},
%	\end{aligned}
%	\label{eq:LASSO}
%\end{equation}
where $\Vert \cdot \Vert_F$ denotes the Frobenius norm,
% $\mathrm{diag}\left( \cdot \right)$ is the operator that generates a diagonal matrix from the input vector, 
$\rho_n$ the noise variance, $\mI$ the identity matrix, and $\lambda$ is a
non-negative regularization constant.

From \eqref{eq:beamformer-signal}, if station $k$ uses beamforming matrix
$\mW_{k} \in \C^{J\times M}$, the correlation matrix becomes
\begin{align*}
	\mR = \mW^\He \mA
\vec{\Sigma}_s \mA^\He \mW + \sigma^2_n \mW^\He \mW,
\end{align*}
where $\mW \in \C^{LJ \times LM} $ is the block diagonal matrix containing the
beamforming vectors. The signal recovery problem \eqref{eq:LASSO} is then
\begin{equation}
	\argmin_{\hat{\bm{\rho}}_s, \hat{\rho}_n}
	\Vert \hat{\mR} - \mW^\He (\mA_g \diag\left( \hat{\bm{\rho}}_s \right) \mA_g^{\He} - \hat{\rho}_n \mI  )\mW \Vert_{F} + \lambda \Vert \hat{\bm{\rho}}_s \Vert_{1}.
	\label{eq:newLASSO}
\end{equation}
This is the optimization used in the performance analysis.

% This is equivalent to \emph{covariance matching} \cite{Ottersten:1998uc} with an
% $\ell_1$-norm regularization to promote sparse solutions.

\section{Performance analysis}
\label{sec:sim}

%This section presents numerical simulations showing the effectiveness of the
%proposed beamforming method.
% The low band antennas are operating at $10$ MHz -- $90$ MHz, and they are
% irregularly spaced, where the minimum and maximum spacings are $2.4$ meters and
% $81.3$ meters respectively. The wavelengths at their observation frequencies are
% between $3$ to $30$ meters. At the upper part of its operating frequency, low
% band antennas also have grating lobes. The summary of the parameters can be
% found in \tref{LOFAR}.
%\begin{table}[t]
%	\footnotesize
%	\renewcommand\arraystretch{1.4}
%	\centering
%	\begin{tabular}{@{} l  c  c  c  c  @{}}
%		\toprule
%		& \multicolumn{2}{c}{Low Band Antennas} & \multicolumn{2}{c}{High Band Antennas} \\
%		\cmidrule(r){2-3}
%		\cmidrule(l){4-5}
% & min & max & min & max  \\ \hline
%Frequency [MHz] &  $10$ & $90$ & $110$ & $250$   \\
%Wavelength [m] & $3$ & $30$ & $1.2$ & $2.7$   \\
%Antenna spacing [m] & $2.4$  & $81.3$ & $5$ & $45$   \\
%Station baselines [m] & $68$ & $1158$ k & $68$  & $1158$ k\\
%\bottomrule
%	\end{tabular}
%	\caption{LOFAR specifications~\cite{VanHaarlem:2013vf}.}
%	\label{tab:LOFAR}
%\end{table}
This section presents numerical simulation, showing the effectiveness of different station beamforms through
randomization and image recovery using sparse signal recovery optimization \eqref{eq:newLASSO}.
%described in Section \ref{sec:imaging_methods}. 

%\begin{equation}
%	\begin{aligned}
%		& \underset{ \hat{\bm{\rho}}_s, \hat{\rho}_n }{\mathrm{argmin}}
%		& &  \Vert \hat{\mR} - \mW^\He (\mA_g \diag\left( \hat{\bm{\rho}}_s \right) \mA_g^{\He} - \hat{\rho}_n \mI  )\mW \Vert_{F} + \lambda \Vert \hat{\bm{\rho}}_s \Vert_{1}.
%	\end{aligned}
%	\label{eq:newLASSO}
%\end{equation}
% \begin{equation}
% 	\begin{aligned}
% 		& \underset{ \hat{\bm{\rho}}_s , \hat{\rho}_n }{\mathrm{minimize}}
% 		& &  \Vert \hat{\mR} - \mW^\He \mA_g \mathrm{diag} \left( \hat{\bm{\rho}}_s \right) \mA_g^{\He} \mW - \hat{\rho}_n \mW^\He\mW  \Vert_{F} \\
% 		& & &  + \lambda \Vert \hat{\bm{\rho}}_s \Vert_{1}.
% 	\end{aligned}
% 	\label{eq:newLASSO}
% \end{equation}

For comparison of the simulation parameters to an actual radio telescope, we give the LOFAR High Band Antenna parameters.
% In particular, we would like to draw attention to the fact that for a large frequency range, both the
% antennas within a station, and the stations with respect to each other, are
% spaced apart more than half of the wavelength of observation.-
Made up of $48$ stations, the minimum and maximum baselines between each is
$68$m and $1158$km respectively. Each station design is regular with $48$
receiving elements spaced $5$m apart, operating at $110$ MHz -- $250$ MHz, which
corresponds to wavelengths between $1.2$m to $2.7$m. This spaces the high band
antennas and stations more than half a wavelength apart, creating grating lobes
(cf. Section \ref{sec:background}).

For simulation we fixed the number of stations $L$ to $4$, the number of
antennas per station $J$ to $12$, and varied number of beams per station $M$.
There were thus $LM(LM-1)/2=4M(4M-1)/2$ cross-correlations and $LM=4M$
autocorrelations.

%For convenience recall there are $L$ antenna stations each with $J$ antennas and $M$ beams sent per station. These
%beams result in $LM(LM-1)/2$
%cross-correlations and $LM$ autocorrelations. %Signal recovery in the random beamforming

% The station layout used in the experiments is shown in \fref{sta-lay}. 
The spacing between antennas within a station was $5$m, and the observation
frequency $200$ MHz, resulting in an antenna spacing of approximately $3.3$
wavelengths. At each simulation run the stations were positioned within a disk of
radius of $60$ wavelengths at random by approximating Poisson disk sampling
using Mitchell's best candidate algorithm \cite{Mitchell:1991vh}.
% \fref{sta-pos} shows a realization of stations using the simulation setting.
%These antenna and station separations are in agreement with the settings of
%LOFAR.
%
%\begin{figure}[b]
%	\centering
%	\begin{subfigure}[b]{0.48\linewidth}
%		\centering
%		\includegraphics[width=0.96\textwidth]{layout-example}
%		\caption{Station layout}
%		\label{fig:sta-lay}
%	\end{subfigure}
%	\begin{subfigure}[b]{0.48\linewidth}
%		\centering
%		\includegraphics[width=\textwidth]{station-example}
%		\caption{Station positions}
%		\label{fig:sta-pos}
%	\end{subfigure}
%	\caption{Station layout and a realization of station positions. Axes in wavelengths.}
%	\label{fig:sta-lay-pos}
%\end{figure}
%
%
This setting produced grating lobes, which does not adversely affect randomized
beamforming. For best-case matched beamforming we attenuated them by rotating the
stations with respect to each other (as done in LOFAR).
%An instance of the averaged beamform over the station beam shapes can be seen
%in \fref{av-db}.
%
%\begin{figure}[b]
%	\centering
%	\includegraphics[width=0.71\linewidth]{db-average-2}
%	\caption{Averaged station response in dB. Grating lobes are seen to be attenuated by rotation of stations.}
%	\label{fig:av-db}
%\end{figure}

% Sources were over the circular region of the $l-m$ plane with radius $0.6$, a
% fairly large field of view. For matched beamforming when $M = 1$, we direct the
% antennas to the center of the region of interest, when $M > 1$, the antennas are
% matched to directions chosen by best candidate sampling to cover large area when
% combined.

Source intensities were chosen from a Rayleigh distribution with second moment
equal to $2$. The sources were contained in a circular region of the $l-m$ plane
with radius $0.6$, a fairly large field of view. The locations were chosen on a
grid of $600$ grid points by the best candidate algorithm to have them spread
over the field of view. For matched beamforming when $M = 1$, we directed the
antennas to the center of the region of interest. When $M > 1$, the antennas were
matched to directions chosen by best candidate sampling to cover large area when
combined.

% Because the performance of sparse signal recovery algorithms degrade with basis
% mismatch~\cite{Chi:2010bk}, the sources were generated on a grid of $600$ grid
% points. This is a fair assumption for comparison. The source positions were also
% chosen by the best candidate algorithm to have them spread over the field of
% view, and the intensities of the are chosen from the Rayleigh distribution with
% second moment equal to $2$.

% Because the performance of sparse signal recovery algorithms degrade with basis
% mismatch~\cite{Chi:2010bk}, the sources were generated. This is a fair
% assumption for comparison.

We evaluated the performance of sky image reconstruction through mean squared
error defined as
\begin{align*}
	\mathrm{MSE} = \frac{1}{ N_\mathrm{I} } \sum_{i = 1}^{ N_\mathrm{I} } \left( \bm\rho_i - \hat{\bm\rho}_i \right)^2,
\end{align*}
where $\bm \rho$ was the true sky image in the vectorized form,
$\hat{\bm{\rho}}$ its estimate, and $N_{ \mathrm{I} }=50$ the number of
simulation runs.

\begin{figure}[t]
	\centering
	\includegraphics[width=0.65\linewidth]{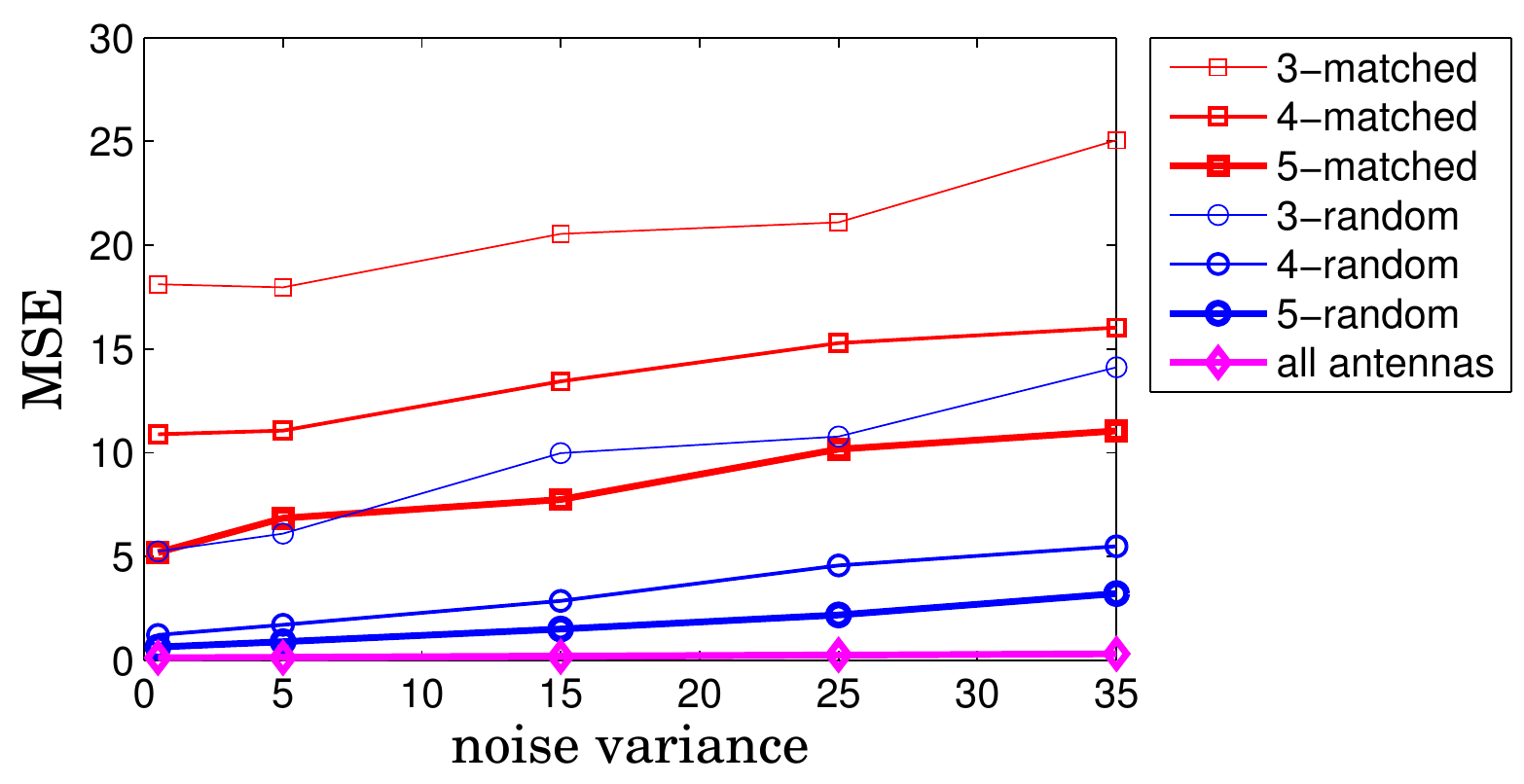}
	\caption{Reconstruction MSE for randomized and matched beamforming for different number of beams.} %at stations.}
	\label{fig:error-plot-noisy}
\end{figure}

\fref{error-plot-noisy} shows the MSE obtained by randomized and conjugate
matched beamforming, as well as a reference result obtained by correlation
directly of every antenna signal pair. The number of beams per station ranged
from $3$ to $5$. The number of sources was fixed to $20$. 
Both methods had similar behavior as the measurement noise varied.
Using three beams per station with randomized beamforming had similar accuracy
to using $5$ with matched beamforming.

\fref{compress} is the dual of \fref{error-plot-noisy}. Noise variance was fixed
while the number of beams per station changed. Except for the case of a single
beam per station, the MSE performance of randomized beamforming was superior to
matched beamforming. The diverse beam shapes generated by randomized beamforming
results in different scaling of sky directions for different station pairs, and
thus, helped resolve sources by also making use of the signal magnitudes. The
diversity of the randomized beam shapes also explains the relative performance
loss when using one beam. When correlating beamformed outputs from two stations,
if low magnitude response directions from one beam overlap with high magnitude
response directions from the other, the effective signal power is attenuated as
can be seen in \eqref{eq:correlation-between-two-beamfors}.
%This can be the case when using
%very small number of beamforms per station for random beamforming. 
Nevertheless, randomized beamforming improved rapidly and outperformed matched
beamforming once there were at least two antennas per station.

\begin{figure}[t]
	\centering
	\includegraphics[width=.65\linewidth]{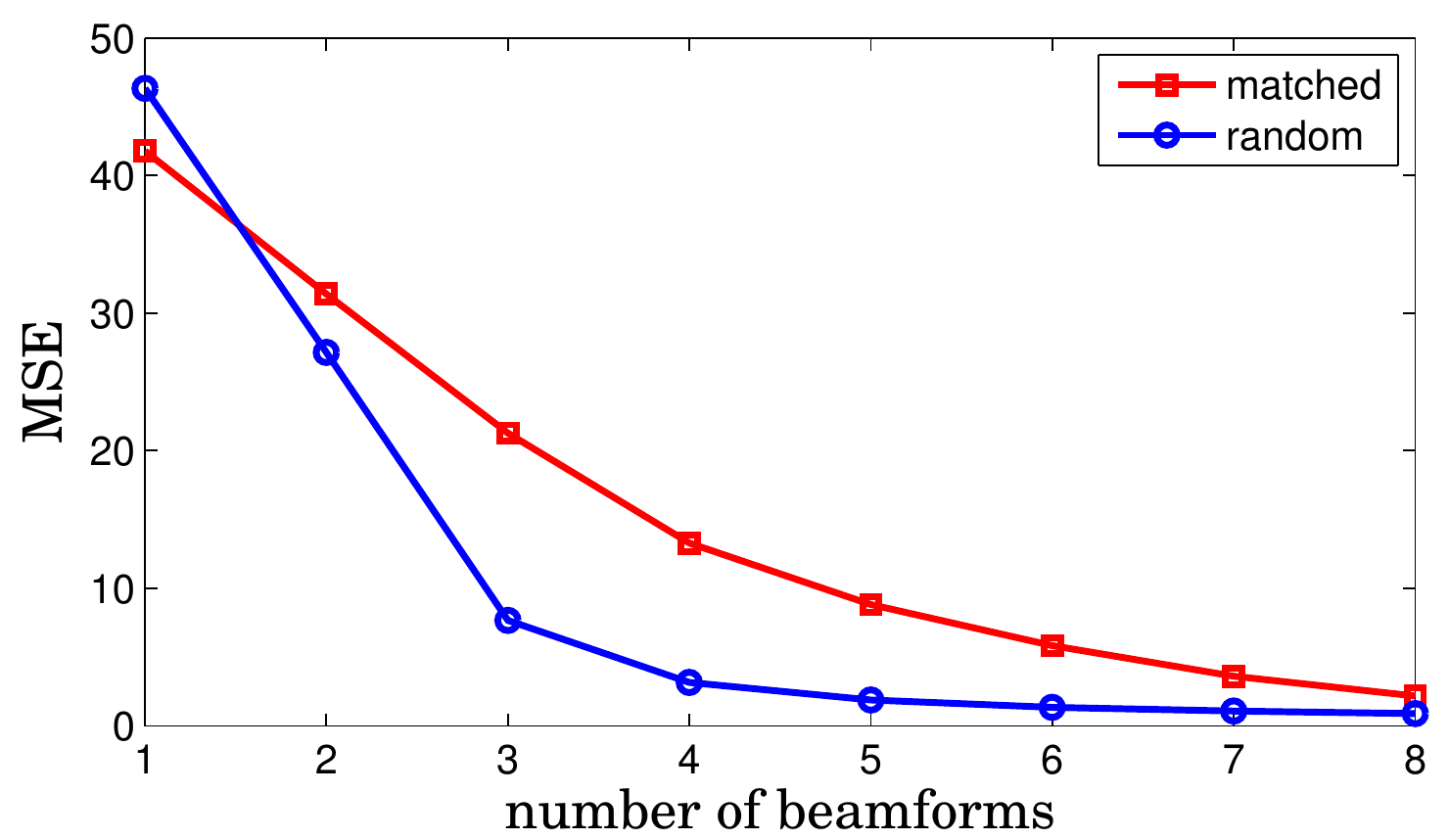}
	\caption{
MSE for randomized and matched beamforming as a function of the number of beams
per station. Noise variance $\sigma^2_n = 25$.}
	\label{fig:compress}
\end{figure}

\begin{table}%[t]
\tbl{Minimum number of beams necessary to meet MSE criteria for matched/randomized
beamforming. Noise variance $\sigma^2_n = 25$.}{
	\footnotesize
	\renewcommand\arraystretch{1.4}
	\centering
	\begin{tabular}{@{} c  c  c  c @{}}
		\toprule
		& \multicolumn{2}{c}{Number of beams} & \\
		\cmidrule(l){2-3}
		MSE  & Matched & Random & Rate ratio \\ \hline
		10   & $5$   & $3$    & 60 \%   \\
		6    & $6$   & $4$    & 67 \%   \\
		3    & $7$   & $4$    & 57 \%   \\
		2    & $8$   & $5$    & 63 \%   \\
		\bottomrule
	\end{tabular}
	\label{tab:MSE}
}
\end{table}

\tref{MSE} shows data compression rates from \fref{compress},
specifically the required number of beams to go below an MSE level when using
randomized and matched beamforming. 
%The column `rate ratio' shows the
%ratio of the data rate by random beamforming to the data rate by matched
%beamforming in order to meet the same MSE criterion. 
As can be seen, the rate reduction for meeting the same MSE can be up to $43 \%$.

\section{Conclusions}
\label{sec:conclusions}

% summary, what we've done

We observed that beamforming in radio interferometry has yet to be fully
exploited, its goals to date somewhat narrow in scope. It seemed attractive to
maximize information in beamforming, getting beams to work in unison. Towards
this end, we introduced a randomized beamforming strategy that increases
measurement diversity. We showed that it can achieve
substantial data reduction while preserving imaging quality. Sparse image
reconstruction, a popular topic in radio astronomy, could additionally benefit
from this ``jumbling up" to a lower dimensional space.

% where do we go from here
We believe the SKA could benefit from a flexible, configurable beamforming
architecture. There is always a resolution and computational trade-off to be
made, and this would ideally be dictated by the science case. To which end, some
of the future work we envisage includes developing further robustness in the
presence of large measurement noise.
%  and adapting optimally the algorithm to the sky as information is gathered. 
Adaptation to other array signal processing problems, such as medical imaging or
seismology, could also be a fruitful avenue of investigation.

\clearpage

% use balance if the list of references becomes a lot
\balance 

\bibliographystyle{ACM-orhan-truncatednames}
\bibliography{skapapers-arxiv}


\begin{thebibliography}{00}

%%% the Web address.

\ifx \showCODEN    \undefined \def \showCODEN     #1{\unskip}     \fi
\ifx \showDOI      \undefined \def \showDOI       #1{{\tt DOI:}\penalty0{#1}\ }
  \fi
\ifx \showISBNx    \undefined \def \showISBNx     #1{\unskip}     \fi
\ifx \showISBNxiii \undefined \def \showISBNxiii  #1{\unskip}     \fi
\ifx \showISSN     \undefined \def \showISSN      #1{\unskip}     \fi
\ifx \showLCCN     \undefined \def \showLCCN      #1{\unskip}     \fi
\ifx \shownote     \undefined \def \shownote      #1{#1}          \fi
\ifx \showarticletitle \undefined \def \showarticletitle #1{#1}   \fi
\ifx \showURL      \undefined \def \showURL       #1{#1}          \fi

\bibitem[\protect\citeauthoryear{Capon}{Capon}{1969}]%
        {Capon:1969tf}
{J Capon}. 1969.
\newblock \showarticletitle{{High-resolution frequency-wavenumber spectrum
  analysis}}.
\newblock {\em Proc. IEEE\/} {57}, 8 (1969), 1408--1418.
\newblock


\bibitem[\protect\citeauthoryear{Dewdney, Hall, Schilizzi, and Lazio}{Dewdney
  et~al\mbox{.}}{2009}]%
        {Dewdney:2009bq}
{P~E Dewdney}, {P~J Hall}, {R~T Schilizzi}, {and} {T~J L~W Lazio}. 2009.
\newblock \showarticletitle{{The Square Kilometre Array}}.
\newblock {\em Proc. IEEE\/} {97}, 8 (Aug. 2009), 1482--1496.
\newblock


\bibitem[\protect\citeauthoryear{Donoho}{Donoho}{2006}]%
        {Donoho:2006vp}
{D~L Donoho}. 2006.
\newblock \showarticletitle{{Compressed sensing}}.
\newblock {\em IEEE Trans. Inf. Theory\/} (April 2006), 1289--1306.
\newblock


\bibitem[\protect\citeauthoryear{Ellingson, Clarke, Cohen, Craig, Kassim,
  Pihlstr{\"o}m, Rickard, and Taylor}{Ellingson et~al\mbox{.}}{2009}]%
        {Ellingson:2009gg}
{S~W Ellingson}, {T~E Clarke}, {A Cohen}, {and} others. 2009.
\newblock \showarticletitle{{The Long Wavelength Array}}.
\newblock {\em Proc. IEEE\/} {97}, 8 (2009), 1421--1430.
\newblock


\bibitem[\protect\citeauthoryear{Jaffer}{Jaffer}{1988}]%
        {Jaffer:1988iw}
{A~G Jaffer}. 1988.
\newblock \showarticletitle{{Maximum likelihood direction finding of stochastic
  sources: a separable solution}}. In {\em IEEE Int. Conf. Acoust., Speech, and
  Signal Proc.} New York, NY, 2893--2896.
\newblock


\bibitem[\protect\citeauthoryear{Lonsdale, Cappallo, Morales, Briggs,
  Benkevitch, Bowman, Bunton, Burns, Corey, deSouza, Doeleman, Derome,
  Deshpande, Gopala, Greenhill, Herne, Hewitt, Kamini, Kasper, Kincaid, Kocz,
  Kowald, Kratzenberg, Kumar, Lynch, Madhavi, Matejek, Mitchell, Morgan,
  Oberoi, Ord, Pathikulangara, Prabu, Rogers, Roshi, Salah, Sault, Shankar,
  Srivani, Stevens, Tingay, Vaccarella, Waterson, Wayth, Webster, Whitney,
  Williams, and Williams}{Lonsdale et~al\mbox{.}}{2009}]%
        {Lonsdale:2009ij}
{C~J Lonsdale}, {R~J Cappallo}, {M~F Morales}, {and} others. 2009.
\newblock \showarticletitle{{The Murchison Widefield Array: Design Overview}}.
\newblock {\em Proc. IEEE\/} {97}, 8 (Aug. 2009), 1497--1506.
\newblock


\bibitem[\protect\citeauthoryear{Mitchell}{Mitchell}{1991}]%
        {Mitchell:1991vh}
{D~P Mitchell}. 1991.
\newblock \showarticletitle{{Spectrally optimal sampling for distribution ray
  tracing}}.
\newblock {\em SIGGRAPH Comput. Graph.\/} {25}, 4 (July 1991), 157--164.
\newblock


\bibitem[\protect\citeauthoryear{Taylor, Carilli, Perley, and {National Radio
  Astronomy Observatory (U.S.)}}{Taylor et~al\mbox{.}}{1999}]%
        {Taylor:1999vv}
{G~B Taylor}, {C~L Carilli}, {R~A Perley}, {and} {{National Radio Astronomy
  Observatory (U.S.)}}. 1999.
\newblock {\em {Synthesis imaging in radio astronomy II}}.
\newblock ASP Conf. Series.
\newblock


\bibitem[\protect\citeauthoryear{Tibshirani}{Tibshirani}{1996}]%
        {Tibshirani:1996wb}
{R Tibshirani}. 1996.
\newblock \showarticletitle{{Regression shrinkage and selection via the
  lasso}}.
\newblock {\em J. Roy. Statist. Soc. Ser. B\/}  {58} (1996), 267--288.
\newblock


\bibitem[\protect\citeauthoryear{van~der Veen and Wijnholds}{van~der Veen and
  Wijnholds}{2013}]%
        {vanderVeen:2013uz}
{A~J van~der Veen} {and} {S~J Wijnholds}. 2013.
\newblock \showarticletitle{{Signal Processing Tools for Radio Astronomy}}.
\newblock Springer New York, New York, NY, 421--463.
\newblock


\bibitem[\protect\citeauthoryear{Van~Haarlem, Wise, Gunst, Heald, McKean,
  Hessels, De~Bruyn, Nijboer, Swinbank, Fallows, Brentjens, Nelles, Beck,
  Falcke, Fender, H{\"o}randel, Koopmans, Mann, Miley, R{\"o}ttgering,
  Stappers, Wijers, Zaroubi, van~den Akker, Alexov, Anderson, Anderson, van
  Ardenne, Arts, Asgekar, Avruch, Batejat, B{\"a}hren, Bell, Bell, van Bemmel,
  Bennema, Bentum, Bernardi, Best, B{\^\i}rzan, Bonafede, Boonstra, Braun,
  Bregman, Breitling, van~de Brink, Broderick, {Broekema, P. C.}, Brouw,
  Br{\"u}ggen, Butcher, van Cappellen, Ciardi, Coenen, Conway, Coolen,
  Corstanje, Damstra, Davies, Deller, Dettmar, van Diepen, Dijkstra, Donker,
  Doorduin, Dromer, Drost, van Duin, Eisl{\"o}ffel, van Enst, Ferrari,
  Frieswijk, Gankema, Garrett, de~Gasperin, Gerbers, de~Geus, Grie{\ss}meier,
  Grit, Gruppen, Hamaker, Hassall, Hoeft, Holties, Horneffer, van~der Horst,
  van Houwelingen, Huijgen, Iacobelli, Intema, Jackson, Jelic, de~Jong, Juette,
  Kant, Karastergiou, Koers, Kollen, Kondratiev, Kooistra, Koopman, Koster,
  Kuniyoshi, Kramer, Kuper, Lambropoulos, Law, van Leeuwen, Lemaitre, Loose,
  Maat, Macario, Markoff, Masters, McKay-Bukowski, Meijering, Meulman, Mevius,
  Middelberg, Millenaar, Miller-Jones, Mohan, Mol, Morawietz, Morganti,
  Mulcahy, Mulder, Munk, Nieuwenhuis, van Nieuwpoort, Noordam, Norden, Noutsos,
  Offringa, Olofsson, Omar, Orr{\'u}, Overeem, Paas, Pandey-Pommier, Pandey,
  Pizzo, Polatidis, Rafferty, Rawlings, Reich, de~Reijer, Reitsma, Renting,
  Riemers, Rol, Romein, Roosjen, Ruiter, Scaife, van~der Schaaf, Scheers,
  Schellart, Schoenmakers, Schoonderbeek, Serylak, Shulevski, Sluman, Smirnov,
  Sobey, Spreeuw, Steinmetz, Sterks, Stiepel, Stuurwold, Tagger, Tang, Tasse,
  Thomas, Thoudam, Toribio, van~der Tol, Usov, van Veelen, van~der Veen, ter
  Veen, Verbiest, Vermeulen, Vermaas, Vocks, Vogt, de~Vos, van~der Wal, van
  Weeren, Weggemans, Weltevrede, White, Wijnholds, Wilhelmsson, Wucknitz,
  Yatawatta, Zarka, Zensus, and van Zwieten}{Van~Haarlem et~al\mbox{.}}{2013}]%
        {VanHaarlem:2013vf}
{M~P Van~Haarlem}, {M~W Wise}, {A~W Gunst}, {and} others. 2013.
\newblock \showarticletitle{{LOFAR: The LOw-Frequency ARray}}.
\newblock {\em Astron. Astrophys.\/}  {556} (Aug. 2013), A2.
\newblock


\bibitem[\protect\citeauthoryear{Wijnholds}{Wijnholds}{2010}]%
        {Wijnholds:2010wn}
{S~J Wijnholds}. 2010.
\newblock {\em {Fish-eye Observing with Phased Array Radio Telescopes}}.
\newblock Ph.D. Dissertation. Delft University of Technology, Delft.
\newblock


\bibitem[\protect\citeauthoryear{Wright}{Wright}{2002}]%
        {Wright:2002th}
{M~C~H Wright}. 2002.
\newblock {\em {A model for the SKA}}.
\newblock {T}echnical {R}eport~16.
\newblock


\end{thebibliography}

\end{document}